\documentclass[10pt,twocolumn,final]{IEEEtran}
\usepackage{lipsum}
\usepackage{amsmath,graphicx}
\usepackage{epstopdf}
\usepackage{verbatim}
\usepackage{mathrsfs}
\usepackage{cite}
\usepackage{amsthm}
\usepackage{epsfig,psfrag}
\usepackage{amsfonts,amssymb}
\usepackage{color}
\usepackage{enumitem}
\usepackage{float}
\usepackage{tikz}
\usepackage{pgfplots}
\pgfplotsset{compat=newest}

\newtheorem{myremark}{Remark}
\newtheorem{theorem}{Theorem}

\newtheorem{defn}{Definition}[section]

\begin{document}

\title{An Algebraic Approach to a Class of Rank-Constrained Semi-Definite Programs With Applications}

\author{Matthew~W.~Morency~\IEEEmembership{Student Member,~IEEE}
	and Sergiy~A.~Vorobyov,~\IEEEmembership{Senior Member,~IEEE}
	\thanks{M.~W.~Morency is with the Dept. Microelectronics, School of Electrical Engineering, Mathematics, and Computer Science, Delft University of Technology, Mekelweg 4, 2628 CD Delft, The Netherlands; email: {\tt M.W.Morency@tudelft.nl}. He was with the Dept. Signal Processing and Acoustics, Aalto University, PO Box 13000 FI-00076, Finland.
	S.~A.~Vorobyov is with the Dept. Signal Processing and Acoustics, Aalto University, PO Box 13000 FI-00076 Aalto, Finland; email: {\tt svor@ieee.org}.
	Preliminary results of this work in application to the problem of rank-constrained transmit beamspace design in multiple-input multiple-output radar have been presented at the {\it IEEE CAMSAP 2015} and have been awarded the first prize Best Student Paper Award. 
	S.~A.~Vorobyov is the corresponding author.
	}%
}

\maketitle

\begin{abstract}
\label{sec:Abstract}
A new approach to solving a class of rank-constrained semi-definite programming (SDP) problems, which appear in many signal processing applications such as transmit beamspace design in multiple-input multiple-output (MIMO) radar, downlink beamforming design in MIMO communications, generalized sidelobe canceller design, phase retrieval, etc., is presented. The essence of the approach is the use of underlying algebraic structure enforced in such problems by other practical constraints such as, for example, null shaping constraint. According to this approach, instead of relaxing the non-convex rank-constrained SDP problem to a feasible set of positive semidefinite matrices, we restrict it to a space of polynomials whose dimension is equal to the desired rank. The resulting optimization problem is then convex as its solution is required to be full rank, and can be efficiently and exactly solved. A simple matrix decomposition is needed to recover the solution of the original problem from the solution of the restricted one. We show how this approach can be applied to solving some important signal processing problems that contain null-shaping constraints. As a byproduct of our study, the conjugacy of beamfoming and parameter estimation problems leads us to formulation of a new and rigorous criterion for signal/noise subspace identification. Simulation results are performed for the problem of rank-constrained beamforming design and show an exact agreement of the solution with the proposed algebraic structure, as well as significant performance improvements in terms of sidelobe suppression compared to the existing methods.     
\end{abstract}


\section{Introduction}
\label{sec:Intro}
According to the signal space concept, a signal is represented as a vector in a vector space \cite{Wozencraft}, \cite{VanTrees}. Working in vector spaces is yet the most popular approach while designing signal processing models and algorithms. As a consequence, much of signal processing is written in the language of linear algebra. That is, the algebra that deals with vectors, and vector spaces, and whose operations are vector addition and composition of linear maps. This language has two appealing features. First, many problems are well described or approximated by linear models. Second, linear algebra is well behaved in the sense that most questions which can be posed within linear algebra can generally be easily answered within this framework. 

As a further development of signal processing toward the use of advanced algebra, frame theory has found fruitful applications for efficient signal/image representation and filter-bank design \cite{Frames}, \cite{Mallat}. More recently, graphical signal processing (GSP) has been introduced for addressing the need of designing signal processing methods for large-scale social networks, big data or biological networks analysis \cite{Shuman}--\cite{Moura2}. Since signals in GSP are defined over a discrete domain whose elementary units (vertices) are related to each other through a graph, algebra beyond linear algebra is gaining more and more attention. In fact, new applications demonstrate that not all questions that are of interest to signal processing engineers can be readily solved by the tools of linear algebra, even if they have their genesis in linear models. 

To this end, some classic and more recent signal processing problems that have been addressed within a linear algebra framework can be more accurately, efficiently, and more completely addressed with the use of advanced algebraic tools, which are not yet widely known and used within signal processing. One example of such problem is the matrix completion \cite{MC}, \cite{Franz1}, which has received an enormous amount of attention, especially since the advent of the NetFlix prize, and the relevant phase retrieval problem \cite{VoroCandes}, \cite{Franz2}. Interestingly, the formulation of the matrix completion problem has much in common with some other important problems in signal processing. To name a few, optimal downlink beamforming design in  multiple-input multiple-output (MIMO) wireless communications \cite{BengtOtt}, \cite{HuangPalomar}; robust adaptive beamforming design for general-rank signal model \cite{ArashSergiy1}; transmit beamspace design in MIMO radar \cite{Fuhrmann}--\cite{ArashJournal}; interference alignment \cite{Dimakis}; and even the classic generalized sidelobe canceller (GSC) design \cite{Frost}, \cite{GriffithsJim} are all related to each other in the sense that they can be formulated as rank-constrained semi-definite programming (SDP) problems. The rank constraint makes such problems non-convex and in general very difficult to solve. 

In the case of the rank-one constraint, SDP relaxation and randomization techniques have been used to address such problems following the landmark work of Goemans and Williamson \cite{GW}. That is, the rank-one constraint can be dropped, and then the resulting convex SDP problem can be solved and a rank-one solution recovered through a randomization procedure. However, the guarantees proven for the problem of finding maximum cut in a graph in \cite{GW} (see also for more generic exposition and other applications \cite{TomMagazin}, \cite{KevinMe}) apply to rank-one constrained problems, while similar guarantees have not, to our knowledge, been shown to hold for general rank constrained problems. However, it may not be necessary to rely on relaxation in order to solve some rank-constrained SDP problems of a high practical importance. Indeed, it is typical, at least it is the case in many of the aforementioned problems, that a strong underlying algebraic structure is enforced by design requirements and other constraints in the problems. Such underlying algebraic structures allow one to significantly simplify the problems.

In this paper,\footnote{Preliminary results for the problem of rank-constrained transmit beamspace design for MIMO radar have been reported in \cite{CAMSAP15}.} the general rank constraint in a class of rank-constrained SDP problems is made implicit by restricting the dimension of the problem to be equal to the desired rank, while requiring the solution to be full-rank. Specifically, we develop an approach by which the dimension of the rank-constrained SDP problems can be restricted by exploiting the underlying polynomial structure of the problems. In this respect, advanced algebra (beyond linear algebra) is instrumental and allows to solve a class of rank-constrained SDP problems in a space of polynomials. The space of polynomials has desirable properties by its definition and forms a feasible and convex restriction of the original optimization problem. Then the obtained SDP problem will be convex and can be efficiently solved. An additional advantage is a lower dimension of the resulting problem that is equal to the rank required by the rank constraint. In this way, many aforementioned problems can be very efficiently solved.

The rest of the paper is organized as follows. Section~II defines and explores some important algebraic preliminaries such as groups, subgroups, rings, and ideals, which are instrumental in developing our approach. The considered type of rank-constrained SDP problems is introduced in Section~III, and a basic motivating result is proved. Section~IV revises some classical signal processing problems such as phase retrieval, downlink beamforming design for MIMO communications, and transmit beamspace design for MIMO radar, and shows that the corresponding optimization problems perfectly fit to the structure of the considered type of rank-constrained SDP problems. The underlying algebraic structure enforced in the considered type of rank-constrained SDP problems by null-shaping constraint is investigated in details in Section~V. In Section~VI, it is also shown that the celebrated GSC is a special case of the addressed type of rank-constrained SDP problems and some generalizations of GSC are shown. Finally, the reformulation of such rank-constrained SDP problem into corresponding convex problem is given in Section~VII, and algebraic consequences of the conjugacy of beamforming and parameter estimation problems are discussed in Section~VIII within the insightful framework of this paper. Our simulation results for the rank-constrained beamforming design are given in Section~IX, followed by conclusions in Section~X.  

\section{Algebraic Preliminaries}
This paper makes key use of several algebraic structures which are not widely used within the engineering literature. An overview of these structures, their corresponding notations, and relationships between these structures is thus included to aid in the understanding of the content of the paper \cite{Barvinok}--\cite{IVA}.

\begin{defn}
	A group is a set of elements $\cal{G}$ with a binary operation $\bullet$ with the following properties:
	\begin{enumerate}[label=Property \arabic*.,itemindent=*]
		\item $a \bullet b \in \cal{G}$,\ $\forall a,b \in \cal{G}$
		\item $\exists e \in \cal{G}\ |$ $\ a \bullet e = e \bullet a = a,\ \forall a \in \cal{G}$
		\item $\forall a \in \cal{G}$,\ $\exists b \in \cal{G} |$ $a \bullet b = b \bullet a = e$
		\item $(a \bullet b)\bullet c = a\bullet(b \bullet c)$
	\end{enumerate}\label{def:group}
\end{defn}
If a group $G$ is commutative with respect to $\bullet$ then G is said to be Abelian.  As an example, consider the set of all permutations of an $n$-tuple. This set is a group (in fact, it is known as the symmetric group, $S_n$), with operation of composition of functions.\footnote{Note, that although it is common to refer to the operation $\bullet$ as multiplication or addition, the operation $\bullet$ need not be the conventional notion of multiplication or addition.} In this case, the identity element is the zero permutation, which takes each element of the $n$-tuple to itself. This is not an Abelian group as composition of functions depends, generally speaking, on the order of the composition. An example of an Abelian group are the integers under addition. 

\begin{myremark}
Every vector space is an Abelian group with respect to vector addition, which is a property which will be used later.
\end{myremark}
Building on this definition, we next introduce the concept of a subgroup.

\begin{defn}
	A subgroup $H < G$ is a subset of $G$ for which all of the group properties hold with respect to $\bullet$.
\end{defn}
Thus, every subgroup is a group unto itself which is also contained in $\cal{G}$. A subgroup of $\cal{G}$ which does not contain every element of $G$ is said to be a proper subgroup of $\cal{G}$. A subgroup of an Abelian group is Abelian by definition. 

\begin{myremark}
Since all vector spaces are Abelian groups with respect to vector addition, it follows that every subspace of a vector space is an Abelian subgroup of that space with respect to addition.
\end{myremark}

Next we define a ring, which is then will be used for defining an ideal and then polynomial ideal. Polynomial deals, in turn, will be used in the paper for developing our approach.

\begin{defn}
	A ring is a set $\cal{R}$ with operations $\bullet$  and $\diamond$ with the following properties: 
	\begin{enumerate}[label=Property \arabic*.,itemindent=*]
		\item $(\cal{R},\bullet)$ is an Abelian group
		\item $\exists 1 \in \cal{R}\ |$ $\ 1\diamond r = r \diamond 1 = r,\ \forall r \in \cal{R}$
		\item $(a \diamond b) \diamond c = a \diamond (b \diamond c), \forall a,b,c \in \cal{R}$
		\item $a\diamond(b \bullet c) = (a\diamond b)\ \bullet (a\diamond c), \forall a,b,c \in \cal{R}$
		\item $(a \bullet b) \diamond c = (a\diamond c) \bullet (b\diamond c), \forall a,b,c \in \cal{R}$
	\end{enumerate}	\label{def:ring}
\end{defn}
Properties $2$ and $3$ of Definition~\ref{def:ring} taken together mean that a ring $\cal{R}$ is a \textit{monoid} under multiplication $\diamond$ operation. Properties $4$ and $5$ of Definition~\ref{def:ring} again taken together simply state that in a ring, multiplication $\diamond$ is left and right distributive over addition $\bullet$. We will only consider commutative rings which have the addition property that $a \diamond b = b\diamond a, \forall a,b \in R$.\footnote{A ring with the property that $R\setminus \{0\}$ is an Abelian group with respect to multiplication $\diamond$ is a Field (e.g. the set of real numbers).} 

The major substructures of rings are known as ideals.

\begin{defn}
	An ideal $\mathcal{I}$ in a commutative ring $R$ is a subgroup of $R$ with the following properties:
	\begin{enumerate}[label=Property \arabic*.,itemindent=*]
		\item $\mathcal{I}$ is a subgroup of R
		\item $\forall a \in \mathcal{I},\ r \in R, a \diamond r \in \mathcal{I}, r \diamond a \in \mathcal{I}$
	\end{enumerate}	\label{def:ideal}
\end{defn}
A relevant example of a ring with a non-trivial ideal, which will be used in the contribution of this paper, is the {\it ring of univariate polynomials over a field $\mathbb{K}$}, which we will denote as $\mathbb{K}[x]$ (read as ``K adjoin x''). To show that this is a commutative ring we need to show that the relevant properties of Definitions~\ref{def:group} and~\ref{def:ring} hold. 

The fact that the polynomials are an Abelian group under addition is obvious since $P(x) + Q(x) = Q(x) + P(x) = R(x) \in \mathbb{K}[x]$, $0 \in \mathbb{K}[x]$, and the additive inverse of a polynomial $P(x)$ is trivially $-P(x) \in \mathbb{K}[x]$. 

To show that it is commutative ring, we note that $A(x)B(x) = B(x)A(x) = C(x) \in \mathbb{K}[x]$ for any polynomials $A(x), B(x), C(x)$ with coefficients in $\mathbb{K}$. The set $\mathbb{K}[x]$ is associative with respect to both multiplication and addition, and polynomial multiplication is distributive over polynomial addition since $A(x)(B(x) + C(x)) = A(x)B(x) + A(x)C(x)$ and $(A(x) + B(x))C(x) = A(x)C(x) + B(x)C(x)$. Finally, we note that the set $\mathbb{K}[x]$ has the multiplicative identity $1$, thus completing the proof.

\begin{figure}[h!]
	\centering
	\centerline{\includegraphics[width=5.9cm]{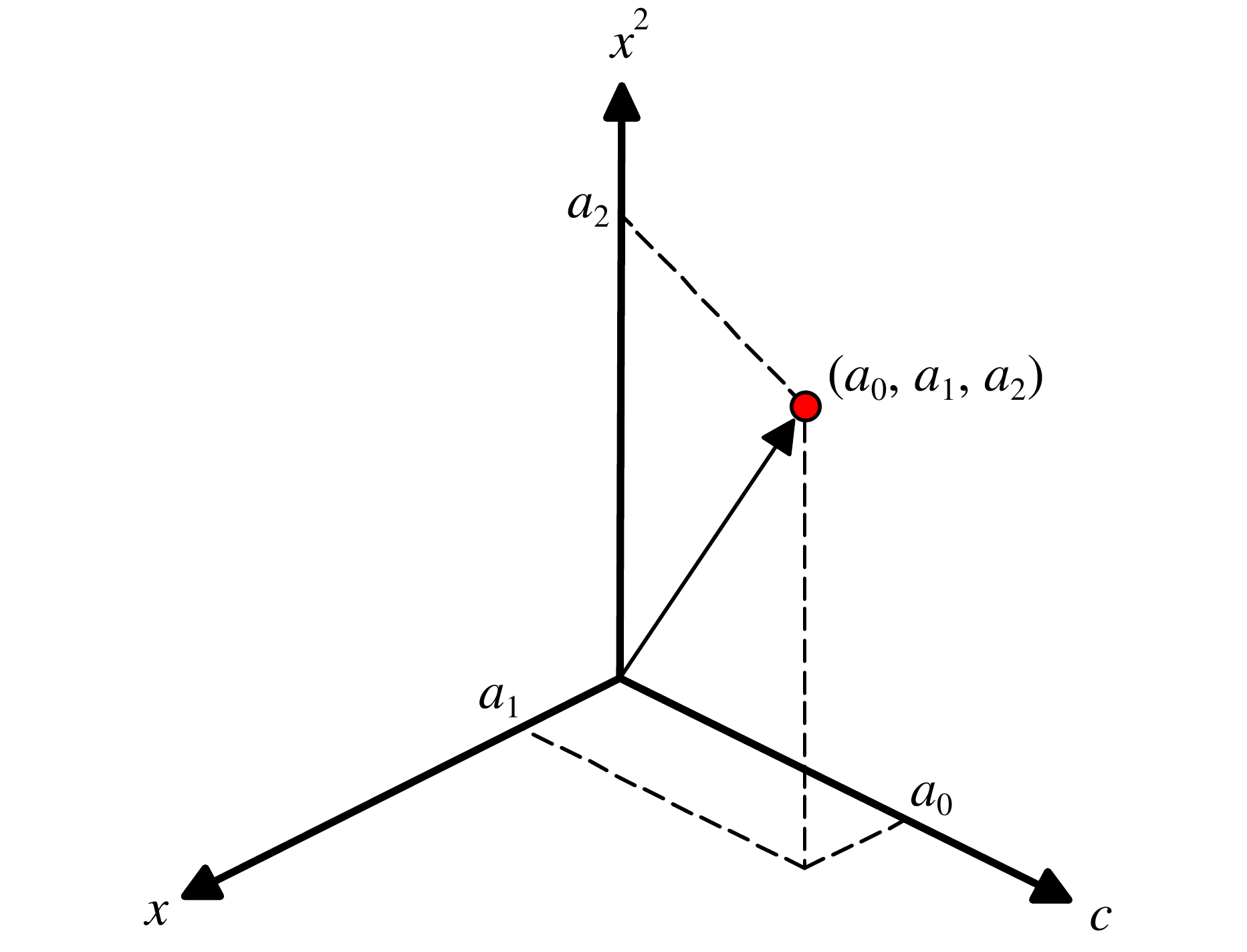}}
	\caption{Depiction of an element of $\mathbb{C}_3[x]$. Notice that a polynomial of degree 2, corresponds to a 3 dimensional vector. Hence, $\mathbb{C}_3[x]$ denotes the space of all polynomials of degree strictly less than 3.}
	\label{fi:coeffspace}
\end{figure}

It is important to mention that polynomials in $\mathbb{K}[x]$ also form a vector space over $\mathbb{K}$. Without restriction on the degree of the polynomials, this space has infinite dimension, and as such a finite dimensional vector space of polynomials implies a restriction of the degree of the polynomials to a finite number $N$. This vector space has a basis of $\{x^i,0\leq i \leq N\}$. We denote the space of polynomials with degree strictly less than $N$ by $\mathbb{K}_{N}[x]$. Fig.~\ref{fi:coeffspace} demonstrates that the subscript $N$ corresponds to the dimension of the vector space, where $N-1$ is the restriction on the degree of the polynomials. Thus, there should be no confusion with the definition of $\mathbb{K}_N[x]$ being the space of polynomials of degree strictly less than $N$. Hereafter in the paper, we consider the case $\mathbb{K} = \mathbb{C}$.

Building on Remark~$2$ and the definition of an ideal (Definition~\ref{def:ideal}), the question arises of when a subset of elements in $\mathbb{C}[x]$ forms an ideal. Let $\mathcal{I}$ be the set of all univariate polynomials $\mathbb{C}[x]$ with a root at $x_0 \in \mathbb C$. This set forms an ideal in the ring $\mathbb{C}[x]$, which we refer hereafter as {\it polynomial ideal}. To see this, consider a polynomial with a single root at $x_0$. By Euclid's division algorithm, a univariate polynomial has a root at a point $x_0$ if and only if it can be written as $P(x) = Q(x)(x - x_1)$. Consider Definition~\ref{def:ideal} and let $P(x) = Q(x)(x - x_1)$. Then $P(x)R(x) = Q(x)R(x)(x-x_1)$ which is again in $\mathcal{I}$. The polynomials with a root at $x_0$ are also clearly a subgroup of $\mathbb{C}[x]$ since $P_1(x) - P_2(x) \in \mathcal I$, where $P_1(x) \triangleq Q_1(x)(x-x_0)$, $P_2(x) \triangleq Q_2(x)(x-x_0)$ for any $Q_1(x),Q_2(x) \in \mathbb{C}[x]$. This ideal is also an infinite dimensional vector space over $\mathbb{C}$. Let $A(x) = P(x)(x-x_1)$ and $B(x) = Q(x)(x-x_1)$. It is easy to show that
\begin{align}
\alpha(A(x) + B(x)) &= \alpha A(x)+\alpha B(x), \alpha \in \mathbb{C}, A(x),B(x) \in \mathcal{I} \nonumber \\ 
&= \alpha(P(x) + Q(x))(x-x_1) \nonumber \\
&= \alpha R(x)(x-x_1),\ R(x) = P(x) + Q(x) \nonumber
\end{align}

Furthermore, it is easy to show that if the polynomials $A(x)$ and $B(x)$ are coprime, that is, if their greatest common divisor (gcd) is a constant, the ideal is the entire ring. To show this, we invoke Bezout's identity
\begin{align}
ap + bq &= gcd(p,q) \nonumber
\end{align}
for some $a,b \in \cal{R}$ where $\cal{R}$ is a principal ideal domain (every ideal is generated by a single element). Since every univariate polynomial ideal is generated by a single element $f \in \mathcal{I}$ \cite{Vinberg}, \cite{IVA}, the univariate polynomials are a principal ideal domain. Moreover, if $1 \in \mathcal{I}$, then $\mathcal{I}$ is the whole ring since $r\diamond 1, 1 \diamond r \in \mathcal{I}, \forall r\in R$. Thus, the only non-trivial subspaces which form an ideal in $\mathbb{C}[x]$ are the subspaces of polynomials with roots in common.

Since every ideal in $\mathbb{C}[x]$ is principal, i.e., can be generated by a single element $f \in \mathbb{C}[x]$, we denote $\mathcal{I} \subset \mathbb{C}[x]$ as$\langle f \rangle \triangleq \{ f \cdot g \ \forall g \in \mathbb{C}[x]\}$, read as ``the ideal generated by'' $f$. The generator of an ideal $\mathcal{I} \subset \mathbb{C}[x]$ is given straightforwardly as $f = \prod_{k} (x - x_k)$. The restriction of this ideal to polynomials with degree less than $N$ is written as $\langle f \rangle |_N$, which is a subspace of $\mathbb{C}^N$. To clarify the notations, it is important to note that $\mathcal{I}|_N \subset \mathbb{C}[x]$ is an algebraic object as a subset of the univariate polynomial ring, while $\langle f \rangle |_N$ is a vector subspace of $\mathbb{C}^N$. The key difference between the two is the notion of scalar multiplication under which both objects are closed. Scalars for the former are elements of the polynomial ring itself, while scalars for the latter are elements of $\mathbb{C}$. The structure which unifies these two concepts is known as a {\it module}, but it is not defined or used here as it is outside of the scope of the paper.

\section{The Problem}
\label{sec:Problem Formulation}
We are interested in solving the following homogeneous quadratically constrained quadratic programming problem
\begin{subequations}
\begin{align}
\min_{\mathbf {W}}\  &\;\mathrm{tr} \{ \mathbf {W}^H \mathbf {C} \mathbf{W} \} \label{InitialProblema} \\
\mathrm{s.t.}      &\;\mathrm{tr} \{ \mathbf{W}^H \mathbf{B}_j \mathbf{W} \} = \delta_j, \; j \in \mathfrak{J} \label{InitialProblemb}
\end{align}
\end{subequations}
where $\mathbf W \in \mathbb{C}^{N \times K}$ is a matrix (vector when $K=1$) of optimization variables, $\mathbf C \in \mathbb{C}^{N\times N}$ and $\mathbf B_j \in \mathbb{C}^{N\times N}$ are matrices of coefficients, $\delta_j \in \mathbb{R}$ is some problem specification parameter, $\mathfrak{J}$ is some index set, $\mathrm{tr} \{\cdot\}$ denotes the trace of a square matrix, and $( \cdot)^H$ stands for the Hermitian transpose of a vector or matrix. Practical problems which can be formulated in this form are nowadays ubiquitous in signal processing and its applications. 

By introducing the new optimization variable $\mathbf X \triangleq \mathbf W \mathbf W^H$, and using the cyclic property of the trace operator, the problem \eqref{InitialProblema}--\eqref{InitialProblemb} can be equivalently rewritten as the following SDP problem
\begin{subequations}
\begin{align}
\min_{\mathbf X}\  &\;\mathrm{tr} \{ \mathbf X \mathbf C \} \label{eq:SDP_obj} \\
\mathrm{s.t.}      &\;\mathrm{tr} \{\mathbf X \mathbf B_j \} = \delta_j, \; j \in \mathfrak{J} \label{eq:affine} \\
               &\;\mathrm{rk} \{\mathbf X \} = K \label{eq:rank}\\
               &\;\mathbf X \succeq 0 \label{eq:PSD}
\end{align}
\end{subequations}
where $\mathrm{rk} \{\cdot\}$ stands for the rank of a matrix and $\mathbf X \succeq 0$ means that matrix  $\mathbf X$ is positive semi-definite (PSD). Assuming that $\mathbf W$ is required to be full rank,\footnote{This is required in many system design problems. For example, in the transmit beamspace design for MIMO radar problem, which we will describe in detail in the next section, a rank-deficient $\mathbf W$ would correspond to a loss of degrees of freedom in the design of the transmit beampattern.} the constraint \eqref{eq:rank} has to be an equality constraint as it is expressed by the definition of $\mathbf X$ and the constraint \eqref{eq:PSD}. 

The problem \eqref{eq:SDP_obj}--\eqref{eq:PSD} is a rank-constrained SDP problem. It is non-convex as a direct consequence of the non-convex constraint \eqref{eq:rank}. The non-convexity of this constraint extends simply from the geometry of the PSD cone. The interior of the cone is exactly the cone of positive definite matrices \cite{Barvinok}, meaning that all rank-deficient solutions must lie on the boundary. Set boundaries are usually non-convex, a notable exception being an affine half-space. 

The dominant approach to addressing this problem following the work of \cite{GW} has been semidefinite relaxation (SDR). That is, the constraint \eqref{eq:rank} is dropped altogether from the problem, and the resulting convex SDP problem is solved. If the solution to the resulting convex problem, denoted as $\mathbf X^{\star}$, is not rank-$K$, it must then be mapped back to the manifold of rank-$K$ $N \times N$ PSD matrices. Obviously, the solution to the original problem \eqref{eq:SDP_obj}--\eqref{eq:PSD} is contained in the feasible set of the relaxed problem. However, what is sacrificed in the relaxation process is a guarantee that the solution to the relaxed problem will be ``close'' to the solution of the original problem. The optimality bound proven to hold in \cite{GW} applies only to rank-one constrained SDP problems.
Then the rank-one solution, i.e., mapping of the solution of the relaxed SDP problem back to the manifold of rank-one matrices, can be obtained with guarantees proven in \cite{GW} through randomization procedure. Similar guarantees have not been shown to hold for general rank-$K$ constrained SDP problems. 

In the case when the problem \eqref{eq:SDP_obj}--\eqref{eq:PSD} is rank-$K$ constrained, but separable, i.e., can be separated to $K$ rank-one constrained SDP sub-problems as in \cite{HuangPalomar}, the optimality bounds proven in \cite{GW} can be directly applied to each resulting sub-problem. However, separating the problem may not be necessary, desirable, or even possible in practice. In such cases, an alternative approach, which we present in this paper, is of importance. 

In contrast to the SDR approach, our approach is based on restriction. That is, rather than expand the feasible set of the optimization problem, we restrict it to lie on the boundary of the cone of PSD matrices. Since the boundary of PSD cone is non-convex, it is not at all obvious that this would result in a convex problem. It is also not obvious how to perform such a restriction. Thus, some care should be taken in with respect to how such a restriction can be performed. The following basic theorem gives a condition for the restriction to be convex.

\begin{theorem}
Let $\cal M$ be a $K$ dimensional subspace of $\mathbb{C}^N$, $\mathcal{R}$ is the set of rank-$K$ $N \times N$ matrices, and $\mathcal{X} \triangleq \{\mathbf X \in \mathcal{R} \, \rvert \, \mathcal{C} \{ \mathbf X \} = {\cal M}\}$, where $\mathcal{C} \{ \cdot \}$ denotes the column space of a matrix. Then the intersection $\mathcal{X} \cap \mathbb{S}_N^+$, where $\mathbb{S}_N^+$ is the cone of $N \times N$ PSD matrices, is convex.
\end{theorem}

{\it Proof:} We must first define a general element in the intersection $\mathcal{X} \cap \mathbb{S}_N^+$. Consider the matrix $\mathbf X \triangleq \mathbf W \mathbf W^H$ where $\mathcal{C} \{ \mathbf W \} = {\cal M}$ and $\mathbf W \in \mathbb{C}^{N \times K}$. The matrix $\mathbf X$ is clearly in $\mathcal{X}$ since $\mathcal{C} \{ \mathbf X \} = {\cal M}$, and also in $\mathbb{S}_N^+$ by construction. Conversely, any element in $\mathcal{X} \cap \mathbb{S}_N^+$ must have this form, since every element in $\mathcal{R}$ must have a factorization of $\mathbf X = \mathbf A \mathbf B^H$ where $\mathbf A$ and $\mathbf B$ are both $N \times K$ matrices, and any matrix in $\mathbb{S}_N^+$ must be Hermitian, implying that $\mathbf A = \mathbf B$.

Now consider two similarly defined matrices in the intersection $\mathcal{X} \cap \mathcal{R}$, denoted as $\mathbf X_1 \triangleq  \mathbf W_1 \mathbf W_1^H$ and $\mathbf X_2 \triangleq \mathbf W_2 \mathbf W_2^H$. Since the column space of $\mathbf W_1$ is the same as column space of  $\mathbf W_2$, i.e., $\mathcal{C} \{ \mathbf W_1 \} = \mathcal{C} \{ \mathbf W_2 \}$, there exists an invertible matrix $\mathbf P$ such that $\mathbf W_2 = \mathbf W_1 \mathbf P$. Taking the convex combination of $\mathbf X_1$ and $\mathbf X_2$, it is easy to see that the train of the following equalities holds
\begin{align}
\lambda \mathbf X_1 + (1 - \lambda) \mathbf X_2 &=  \lambda \mathbf W_1 \mathbf W_1^H + (1 - \lambda) \mathbf W_2 \mathbf W_2^H \nonumber \\ 
&= \lambda \mathbf W_1 \mathbf W_1^H + (1 - \lambda)\mathbf W_1 \mathbf P \mathbf P^H \mathbf W_1^H \nonumber \\
&= \mathbf W_1 \big(\lambda \mathbf I + (1 - \lambda) \mathbf P \mathbf P^H \big)\mathbf W_1^H \nonumber
\end{align}
for $\lambda \in [0,1]$. 

The matrix $\mathbf P \mathbf P^H$ is PSD by construction, and specifically, it is positive definite (PD), since $\mathbf P$ was invertible to begin with. The PD matrices form a cone, and thus, $\mathbf P' = \mathbf I + (1 - \lambda) \mathbf P \mathbf P^H$ is also PD, admitting a further factorization as $\mathbf P' = \mathbf R \mathbf R^H$, with $\mathbf R$ invertible. Hence, $\mathbf W_1 \mathbf P' \mathbf W_1^H \in \mathcal{X} \cap \mathcal{R}$. This complets the proof. $\blacksquare$ 

It is worth observing that the intersection $\mathcal{X} \cap \mathbb{S}_N^+$ introduced in Theorem~$1$ is isomorphic to the cone of PSD matrices of size $K$, and thus, it has dimension $K(K+1)/2$. Therefore, the set of all matrices of the form $\mathbf W \mathbf P \mathbf W^H$ form a $K(K+1)/2$ dimensional face of the PSD cone. Indeed, it can be shown that a set is a face of the PSD cone if and only if it satisfies the parametrization given in Theorem~$1$ \cite{Barvinok}, \cite{ParriloPermenter}.

Summarizing, if one can glean the subspace $\cal V$ which contains the column space of $\mathbf X$ which satisfy the affine constraints in the problem \eqref{eq:SDP_obj}--\eqref{eq:PSD}, among others, Theorem~$1$ provides a clear and direct method to restrict the problem to a smaller dimension. Such procedure of restricting the problem to a smaller dimension contains three sept. First, it is needed either estimate or know the subspace $\cal V$, then produce a basis for it of the appropriate dimension, and finally take this basis to be the columns of $\mathbf W$. 

Section~\ref{sec:Null Shape} provides a complete example of the solution for the subspace $\cal V$ (fist step of the procedure highlighted above) defined by the constraints of a rank-constrained SDP incling the null-shaping constraint, and Section~\ref{sec:dimreduce} shows how, by way of understanding the algebraic structures laid out in the preliminaries, we can arbitrarily restrict the dimension of this space to coincide with the desired rank of the solution, $K$. However, any other method which produces a basis for the space $\cal V$ is compatible with the findings of this paper. Before proceeding to such studies, we would like to stress on the importance and ubiquitous nature of the problem \eqref{eq:SDP_obj}--\eqref{eq:PSD} by explaining some currently important applications in terms of the problem \eqref{eq:SDP_obj}--\eqref{eq:PSD}.

\section{Applications}
\label{sec:Applications}
Many classical and currently important signal processing problems can be expressed in the form \eqref{eq:SDP_obj}--\eqref{eq:PSD}. Such most popular recently problems are listed here, as well as it is shown how they can be formulated in the form \eqref{eq:SDP_obj}--\eqref{eq:PSD}.

\subsection{Phase Retrieval}
\label{subsec:Phase Retrieval}
The simplest form of the objective function \eqref{eq:SDP_obj} is when $\mathbf C = \mathbf I_N$, in which case the objective function becomes $\mathrm{tr} \{\mathbf X \}$. For symmetric (and Hermitian) matrices, this is equivalent to the nuclear norm, and nowadays widely used as a proximal operator for rank reduction. 

For example, consider the problem of phase retrieval \cite{VoroCandes}, in which the goal is to retrieve a vector $\mathbf x_0 \in \mathbb{C}^N$ about which we only have quadratic measurements $\lvert \mathbf a_j^H \mathbf x_0 \rvert^2 = b_j$ where $\mathbf a_j \in \mathbb{C}^N$ and $b_j \in \mathbb {R}$. It means that we observe the intensity measurement $b_j$, while $\mathbf a_j$ is a sampling vector. Simple manipulations using the trace operator show that $\lvert \mathbf a_j, \mathbf x_0 \rvert^2 = \mathrm{tr} \{ \mathbf a_j^H \mathbf x_0 \mathbf x_0^H \mathbf a_j \} = \mathrm{tr} \{ \mathbf X \mathbf A_j \}$ where $\mathbf A_j \triangleq \mathbf a_j \mathbf a_j^H$ and $\mathbf X \triangleq \mathbf x_0 \mathbf x_0^H$. Hence the phase retrieval problem can be equivalently rewritten as the following feasibility problem
\begin{subequations}
\begin{align}
\mathrm{find}\ \ \ \ &\mathbf X \label{eq:feas_obj}\\
\mathrm{s.t.} \ \ \ &\mathrm{tr} \{ \mathbf X \mathbf A_j \} = b_j,\ \forall \ j \\
&\mathrm{rk} \{\mathbf X \} = 1  \\
&\mathbf X \succeq 0. \label{eq:feasSDP}
\end{align}
\end{subequations}

It is easy to check that the problem \eqref{eq:feas_obj}--\eqref{eq:feasSDP} differs from the problem \eqref{eq:SDP_obj}--\eqref{eq:PSD} only in the objective function. However, the difficulty with the problems \eqref{eq:SDP_obj}--\eqref{eq:PSD} and \eqref{eq:feas_obj}--\eqref{eq:feasSDP} lies not with the objective function, but with the feasibility sets, which are of identical form. If the feasibility set of \eqref{eq:feas_obj}--\eqref{eq:feasSDP} is non-empty then it is relatively easy to minimize a linear function over it, and thus, define a problem in the form of \eqref{eq:SDP_obj}--\eqref{eq:PSD}. Therefore, insights about one problem provide insights about the other.

\subsection{Optimal Downlink Beamforming}
\label{subsec:OBP}
Another example of \eqref{eq:SDP_obj}--\eqref{eq:PSD} is the optimal downlink beamforming problem in MIMO wireless communication systems. In this problem, the objective is to minimize the total transmitted power while maintaining an acceptable quality of service for all users. Assuming constant modulus waveforms and total number of $J$ users, the total transmit power is given as $\sum_{j=1}^{J} \|\mathbf w_j\|$ where $\mathbf w_j \in \mathbb{C}^N$ is the beamforming vector corresponding to the $j$-th user. Then the total transmitted power can be rewritten as 
\begin{align}
\mathrm{tr} & \left\{ \sum_{j=1}^J \mathbf w_j \mathbf w_j^H \right\} = \mathrm{tr} \{ \mathbf W^H \mathbf W \} \nonumber \\
& = \sum_{j=1}^J \mathrm{tr} \left\{ \mathbf w_j \mathbf w_j^H \right\} = \sum_{j=1}^J \mathrm{tr} \{ \mathbf X_j \} \label{TotalP}
\end{align}
where $\mathbf W \triangleq [\mathbf w_1,\cdots,\mathbf w_J]$ has rank $J$ and $\mathbf X_j \triangleq \mathbf w_j \mathbf w_j^H$ is the rank-one matrix.  

Constraints on the quality of service for all users are written in term of the signal-to-interference plus noise ratio in the following form
\begin{align}
\frac{\mathbf w_j^H \mathbf R_j \mathbf w_j}{\sum_{i\neq j} \mathbf w_i^H \mathbf R_{i,j} \mathbf w_i + \sigma_j^2} \geq \gamma_j, \; \forall j \label{QofS}
\end{align}
where $\mathbf R_j$ is the signal auto-correlation matrix corresponding to the $j$-th user, $\mathbf R_{i,j}$ is the cross-correlation matrix corresponding to the interfence between the $i$-th and $j$-th users, $\gamma_j$ is a quality-of-service specification for $j$-th user, and $\sigma_j^2$ is the noise variance corresponding to the $j$-th user's channel, assuming the noise observed on each channel is an independent Gaussian random variable. 

Using \eqref{TotalP} and \eqref{QofS} expressed in terms of $\mathbf X_j$, the optimal downlink beamforming problem can be written as
\begin{subequations}
\begin{align}
\min_{\mathbf X_k} \ & \; \sum_{j=1}^{K}\mathrm{tr} \{\mathbf X_j \} \label{eq:OBP_SDP} \\
\mathrm{s.t.} &\; \mathrm{tr} \{ \mathbf X_j \mathbf R_j \} - \gamma_j \sum_{i\neq j} \mathrm{tr} \{ \mathbf X_i \mathbf R_{i,j} \} = \gamma_j \sigma_j^2,\ \forall \ j \\
&\; \mathrm{rk} \{ \mathbf X_j \} = 1,\ \forall \ j \\
&\; \mathbf X_j \succeq 0,\ \forall \ j \label{eq:OBP_SDPc}
\end{align}	
\end{subequations}
which fits the description of the problem \eqref{eq:SDP_obj}--\eqref{eq:PSD}. 

The problem \eqref{eq:OBP_SDP}--\eqref{eq:OBP_SDPc} is separable as the constraints for every $j$-th user are independent of the constraints for the remaining users, and the objective consists of $J$ summanrd one per each user. Then $J$ rank-one constrained SDP problems can be solved for finding $\mathbf X_j$ for all $J$ users. Moreover, it is clear from Theorem~$1$ that the optimal solution for each separate problem with respect to the corresponding $\mathbf X_j$ must lie in a one-dimensional face of the PSD cone. These one-dimensional faces need not coincide. However, as we will show later, by introducing new constraints, it can happen so that each of these one-dimensional faces must exist within a larger $J(J+1)/2$ face of the cone, thus granting great insight into the optimal solution to the problem, and significant computational savings.

\subsection{Transmit Beamspace in MIMO Radar}
\label{subsec:TBMIMORadar}
One more important application problem of the form \eqref{eq:SDP_obj}--\eqref{eq:PSD} is the transmit beamspace design for MIMO radar \cite{Fuhrmann}--\cite{ArashJournal}. The goal of this optimization problem is to match an ideal transmit beampattern as closely as possible with an achievable one, subject to physical constraints imposed by a transmit antenna array. The ideal beampattern is based on presumed knowledge of the locations of targets of interest. Due to the nature of the radar problem, the most that could be assumed about the target locations would be a prior distribution. If one has knowledge of the prior distribution, it makes sense to transmit energy in regions where targets are likely to be located, with less or no energy transmitted in the regions where targets are unlikely to appear or cannot be located. 

For example, if the prior knowledge is that the targets lie within an angular sector $\Theta = [\theta_1, \theta_2]$, then the optimal strategy would be to transmit all of the available power into this sector, while transmitting as low as possible energy elsewhere \cite{TransmitEnergyFocusing}, \cite{ArashJournal}.

In a MIMO radar system, a linear combination of $K$ orthogonal baseband waveforms $\boldsymbol \psi (t) \triangleq [\psi_1(t), \ldots, \psi_K(t)]^T$ is transmitted from an antenna array (we assume a uniform linear array (ULA) for simplicity) of $N$ elements, where $K < N$, in order to concentrate energy over the desired sector $\Theta$. 

The signal at the transmitter at the time instant $t$ in the direction $\theta$ is given by
\begin{equation}\label{eq:TxSignal}
\mathbf s (t,\theta) = \mathbf a^H(\theta) \mathbf W \boldsymbol \psi(t)
\end{equation}
where $[\mathbf a(\theta)]_n \triangleq e^{j2\pi / \lambda_c (n-1) d_x\sin(\theta)},\ \theta \in [-\pi/2,\pi/2]$ denotes the $n$ ($n \in 1,\cdots,N$) element of the array response vector. 

Using \eqref{eq:TxSignal} and orthogonality of the waveforms, the magnitude of the beampattern in any direction $\theta$ is given by the inner product of $\mathbf s(t,\theta)$ with itself integrated over a pulse duration $T$, that is, 
\begin{eqnarray}
G(\theta) &\!\!\!=\!\!\!& \int_{T} \mathbf s (t,\theta)\mathbf s^H (t,\theta)dt \nonumber\\
&\!\!\!=\!\!\!& \mathbf a^H(\theta)\mathbf W \left(\int_{T} \boldsymbol \psi(t) \boldsymbol \psi^*(t)dt \right) \mathbf W^H \mathbf a(\theta) \nonumber\\
&\!\!\!=\!\!\!& \mathbf a^H(\theta)\mathbf W \mathbf W^H \mathbf a(\theta) = {\mathrm tr} \left\{ \mathbf W \mathbf W^H \mathbf a(\theta) \mathbf a^H(\theta) \right\} \nonumber \\
&\!\!\!=\!\!\!& {\mathrm tr} \{ \mathbf X  \mathbf C \} \label{Beampattern}
\end{eqnarray}
where $\mathbf X \triangleq \mathbf W \mathbf W^H$ and $\mathbf C \triangleq {\mathbf a} (\theta) {\mathbf a}^H (\theta)$.

Denoting the desired ideal beampattern by $G_{\mathrm d} (\theta)$ and using \eqref{Beampattern}, the transmit beamspace design problem for MIMO radar can be written as
\begin{subequations}
	\begin{align}
	\min_{\mathbf X} \max_{\theta} &\; \left| G_{\mathrm d} (\theta) - {\mathrm tr} \{ \mathbf X \mathbf C \} \right| \label{eq:opt2}  \\
	\mathrm{s.t.} &\; {\mathrm tr}\{ \mathbf X \mathbf B_j \} = \frac{E}{N},\ j = 1,\cdots,N \label{eq:constraintmatrix} \\
	&\; \mathrm{rk} \{ \mathbf X \} = K \\
	&\; \mathbf X \succeq 0 \label{eq:PSD1}
	\end{align}	
\end{subequations}
where $E$ is the total power budget at the transmit antenna array and $| \cdot |$ stands for the magnitude. 

The constraint \eqref{eq:constraintmatrix} serves to control the distribution of power among antenna elements. The matrix $\mathbf B_j$ is a selection matrix consisting of all zeros except for a $1$ in the $j$-th diagonal position. For example, taking $j = 1$ in \eqref{eq:constraintmatrix} yields $\sum_{k=1}^{K} \left| \mathbf w_{1,k} \right |^2$, which is the amount of power used by all waveforms transmitted from the first antenna. This need not necessarily equal to $E/N$. For instance, certain antennas could be reserved for use in communications in which case one would constrain that antenna to not use any energy for the purposes of transmission of radar signals.

\section{Underlying Algebraic Structure Forced by Null-Shaping Constraint}
\label{sec:Null Shape}
In this section we consider one frequently used constraint, often referred to as null-shaping constraint, which forces a certain algebraic structure in the feasibility set of corresponding optimization problems. Based on this constraint we show how our restriction based approach to solving problems of the form \eqref{eq:SDP_obj}--\eqref{eq:PSD} can be realized. 

As shown above, in several array processing and signal processing applications, the issue of where to not transmit energy can be as important as where to transmit energy. For example, we may know the angular locations of users in adjacent cells of a cellular communication network with whom we wish to not interfere. Another example is for the transmit beamspace-based MIMO radar, where we assume a prior distribution of targets. The strategy is to concentrate energy in the areas of highest probability of target location, and mitigate energy transmission to areas of low (possibly zero) probability of target location. Thus, the null-shaping constraints of the form
\begin{align}
\mathbf a^H(\theta_l)\mathbf W \mathbf W^H \mathbf a(\theta_l) &= 0,\ \; l = 1,\cdots,L. \label{eq:SOS}
\end{align}
are of high practical importance. Here $\theta_l$, $l = 1,\cdots,L$ are the locations in which we do not wish to transmit energy. It can be shown that the vectors $\mathbf w$ which satisfy \eqref{eq:SOS} all lie in the same face of the PSD cone, and thus, a reduction to this face can be easily derived by way of Theorem~$1$. However first, let us consider what equations \eqref{eq:SOS} imply.

Note that equations \eqref{eq:SOS} are sum of squares (SOS) polynomials by construction. Defining $y_k(\theta) \triangleq \mathbf a^H (\theta) \mathbf w_k$, we can rewrite each equation as $\mathbf a^H(\theta_l)\mathbf W \mathbf W^H \mathbf a(\theta_l) = \sum_{k=1}^{K} \left| y_k \right|^2$ from which the SOS nature of \eqref{eq:SOS} is apparent. 

The relation $\sum_{k=1}^{K} \left| y_k \right|^2 = 0$ obviously implies that each $y_k = 0$. Therefore, by introducing the matrix $\mathbf A \triangleq [\mathbf a(\theta_1), \cdots, \mathbf a(\theta_L)]$ and rewriting equations \eqref{eq:SOS} as
\begin{align}
{\mathrm diag}\bigg\{ \mathbf A^H\mathbf W \mathbf W^H \mathbf A \bigg\} &= \mathbf 0 \label{eq:SOS2}
\end{align}
we see that \eqref{eq:SOS} (or equivalently \eqref{eq:SOS2}) can be satisfied if and only if the column space of $\mathbf W$ is a subspace of the nullspace of  $\mathbf A^H$, i.e., $\mathcal{C} \{ \mathbf W\} \subset \mathcal{N} \{\mathbf A^H \} \triangleq \{\mathbf w \in \mathbb{C}^N | \mathbf A^H \mathbf w = \mathbf 0\}$. Here  ${\mathrm diag}\{ \cdot \} $ is the operation that takes the diagonal elements of a square matrix and write them in a vector,  $\mathcal{N} \{\cdot \}$ denotes the nullspace of a matrix, and  $\mathbf 0$ is the vector of all zeros. Equivalently, $\mathbf A$ is in the nullspace of $\mathbf W^H$, however, $\mathbf W$ is the design variable, and thus, we consider $\mathcal{N} \{\mathbf A^H \}$. From the definition of $\mathcal{N} \{\mathbf A^H \}$ it is clear that every $\mathbf w \in \mathcal{N} \{\mathbf A^H \}$ describes the coefficients of a polynomial with roots at the generators of $\mathbf A^H$, denoted as $\alpha_l^*$, that is, 
\begin{align}
\mathbf A^H \mathbf w = \mathbf 0 &\iff \sum_{i=0}^{N-1}(\alpha^*_{l})^i \mathbf w_i = 0,\  \forall l \in 1,\cdots,L . \label{eq:Nulldef1}
\end{align}

A polynomial $P(x)$ has a root at some point $\alpha$ if and only if $(x-\alpha)$ is a factor of $P(x)$ \cite{Vinberg}. By induction, it can be seen that a polynomial $P(x)$ has roots at points $\alpha_1^*, \cdots, \alpha_l^*$ if and only if $P(x) = Q(x)B(x)$ where
\begin{align}
Q(x) \triangleq \prod_{l=1}^{L} (x - \alpha_l^*). \label{eq:Qdef}
\end{align}

From \eqref{eq:Nulldef1} and \eqref{eq:Qdef}, the nullspace $\mathcal{N} \{\mathbf A^H \}$ can be expressed as
\begin{align}
\mathcal{N} \{\mathbf A^H \} &= Q(x) \mathbb{C}_{N-L}[x] = \langle Q(x) \rangle |_{N}  \nonumber 
\end{align}
where $\mathbb{C}_{N-L}[x]$ denotes the space of all polynomials of degree strictly less than $N-L$. The degree is strictly less than $N-L$ as a constant polynomial is defined to have degree $0$. $\mathbb{C}_{N-L}[x]$ has the standard polynomial basis of $\{1,x,x^2,\cdots,x^{N-L-1}\}$, and thus a basis for $\mathcal{N} \{\mathbf A^H \}$ is $\mathcal{B} \triangleq Q(x)\{1,x,x^2,\cdots,x^{N-L-1}\}$. 

Every ideal is first of all an additive Abelian group, as are vector spaces, and thus, the addition of any two elements in the ideal will result in another element in the ideal. Since it is a vector space as well, scaling by an element $\mathbf w$ in $\mathbb{C}$ will yield another element in the ideal. So, if we have a matrix representation of the basis $\mathcal{B}$ denoted by $\mathbf Q$, any matrix of the form $\mathbf Q \mathbf P$ will remain in the ideal $\mathcal {I} |_{N}$. 

Taking $\mathbf P$ to be invertible, we can conclude that, having fixed $L$, the polynomial ideal $\mathcal {I} |_{N}$ describes an entire proper face of the PSD cone of dimension $(N-L)(N-L+1)/2$, by direct comparison with the result of Theorem~$1$. Thus, the inclusion of constraints of the form \eqref{eq:SOS} to any of the problems described in Subsection~\ref{sec:Applications} requires that feasible set is restricted to this face.

\subsection{Construction of $\mathbf Q$}
\label{subsec:ConstQ}
Let $\mathbf q \triangleq [(-1)^{L-1}s_{L-1},(-1)^{L-2}s_{L-2},\cdots,(-1)s_1,1]^T$ where $s_1,\cdots,s_{L-1}$ are the elementary symmetric functions of $\alpha_1^*,\cdots,\alpha_L^*$ and $[ \cdot ]^T$ stands for the transpose. The $k$-th elementary symmetric function in $L$ variables (in this case, $\alpha_1^*,\cdots,\alpha_L^*$) is the sum of the products of the $k$ subsets of those $L$ variables. For example, if $L = 3$ then 
\begin{align}
s_3 &= \alpha_1^* + \alpha_2^* + \alpha_3^* \nonumber \\
s_2 &= (\alpha_1\alpha_2)^* + (\alpha_2\alpha_3)^* + (\alpha_1\alpha_3)^*\nonumber \\
s_1 &= (\alpha_1\alpha_2\alpha_3)^* \nonumber
\end{align}

Let $\mathbf q' \triangleq [\mathbf q,0,\cdots,0]^T\ \in \mathbb{C}^N$. Then a basis of $\mathcal{N} \{ \mathbf A^H \}$ is represented by the columns of the Toeplitz matrix, with $\mathbf q'$ as the first column
\begin{align}
\mathbf Q &= 
\begin{bmatrix}
\includegraphics[scale=0.28]{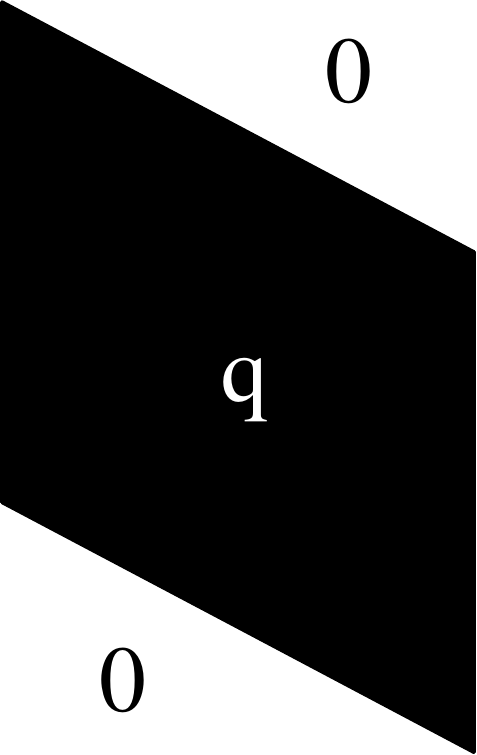} 
\end{bmatrix} . \nonumber 
\end{align}
For a polynomial $Q(x) = a_0 + a_1x + \cdots + a_Lx^L$, with roots $\alpha_1^*,\cdots,\alpha_l^*$, Vi\`ete's formulas yield the coefficients $a_0, \cdots, a_{L-1}$ as
\begin{align}
s_1(\alpha_1^*,\cdots,\alpha_L^*) &= -\frac{a_{L-1}}{a_L} \nonumber \\
s_2(\alpha_1^*,\cdots,\alpha_L^*) &= \frac{a_{L-2}}{a_L} \nonumber \\
&\  \vdots \nonumber \\
s_L(\alpha_1^*,\cdots,\alpha_L^*) &= (-1)^L\frac{a_{0}}{a_L}. \nonumber
\end{align}
Thus, the elements of the vector $\mathbf q$ are the coefficients of $Q(x)$, which are given as a function of the roots of $Q(x)$ by Vi\`ete's formulas, with $a_L = 1$.

\subsection{On Dimensionality Reduction}
\label{sec:dimreduce}
Given a set ${\mathfrak{A}} = \{ \alpha_1, \alpha_2, \cdots, \alpha_L \}$, the previous subsection provides an exact method for construction of the matrix $\mathbf Q$, the columns of which span $\langle Q(x) \rangle |_{N}$. From this construction, it is clear that there is a direct link between the cardinality of $\mathfrak{A}$ and the dimensionality of the column space of $\mathbf Q$.

Specifically, if $|{\mathfrak A}| = L$, then $K \triangleq \dim (\mathcal{C} \{\mathbf Q \}) = N - L$. As such, the construction of $\mathbf Q$ allows for the exact control of the dimension of the problem. That is, given a desired solution of rank $K$, it is possible to use $\mathbf Q$ in tandem with the parametrization given in Theorem~1 to restrict the problem to a $K(K+1)/2$ dimensional face of the PSD cone by choosing $K = N - L$.

Note that $\mathfrak{A}$ is a set on which a collection of polynomials vanish, namely, the ideal $\mathcal{I}$. Such sets are called {\it algebraic varieties}. Formally, the definition of an algebraic variety associated to a particular ideal $\mathcal I$ is given as
\begin{align}
	\mathfrak{V}(\mathcal I) &\triangleq \{ p \in \mathbb{K}^N | f(p) = 0\  \forall f \in \mathcal I \} \nonumber
\end{align}
and an ideal given an associated variety is defined as
\begin{align}
	\mathcal I(\mathfrak{V}) &\triangleq \{ f \in \mathbb{K}[x_1,\cdots,x_n] | f(p) = 0\  \forall p \in \mathfrak{V} \}. \nonumber 
\end{align}

Given a set of varieties and inclusion relationship
\begin{align}
	\mathfrak{V}_0 \subset \mathfrak{V}_1 \subset \cdots \subset \mathfrak{V}_M \nonumber
\end{align}
it is easy to see that taking the ideals associated to these varieties reverses the order of inclusion. Specifically, for the above inclusion relationship
\begin{align}
	\mathcal{I} (\mathfrak{V}_0) \supset \mathcal{I} (\mathfrak{V}_1) \supset \cdots \supset \mathcal{I} (\mathfrak{V}_M) \nonumber
\end{align}
holds \cite{IVA}. For example, take $\mathbb{K}[x_1,\cdots,x_n] = \mathbb{C}[x]$, and let $\mathcal I = \langle (x-1) \rangle$. Clearly, the variety $\mathfrak{V} (\mathcal I) = \{1\}$. Now, consider the variety $\mathfrak{V}' = \{1,2\}$, which clearly contains $\mathfrak{V}$. Its associated ideal $\mathcal{I} (\mathfrak{V}) = \langle (x-1) (x-2) \rangle$ which is obviously contained in $\mathcal{I} (\mathfrak{V})$. Using the definitions provided, it is not difficult to come to a formal proof of the inclusion reversing relationship for general varieties and ideals. This relationship becomes practically important when the number of constraints $L$ of type \eqref{eq:SOS} does not satisfy the relationship $N - L = K$.

Consider, for example in the optimal downlink beamforming problem, the case where a communication system consisting of a transmit array of $N = 20$ antennas at the base station in the cell of interest, $K = 3$ single antenna users in the cell, and $L = 4$ users in adjacent cells at the given directions $\theta_1, \theta_2, \theta_3, \theta_4$. By using constraints \eqref{eq:SOS} corresponding to the directions of the users in the adjacent cells, we constrain the problem dimension to $20-4 = 16$. However, we would like to restrict the problem dimension further to $3$. Using the inclusion reversing relationship between ideals and varieties, it is clear that if $\mathfrak{V}_0 = \{\alpha_1, \alpha_2, \cdots, \alpha_4\}$ and $\mathfrak{V}_M = \{\mathfrak{V}_0, \alpha_{L+1 = 5}, \cdots, \alpha_{M = 17} \}$, then $\mathcal{I} (\mathfrak{V}_M) \subset \mathcal{I} (\mathfrak{V}_0)$.

Thus, the inclusion reversing relationship between ideals and their associated varieties allow us to set roots at arbitrary locations, so long as the new variety $\mathfrak{V}'$ contains the variety described by the equations \eqref{eq:SOS}. In this way, it is possible to both restrict the dimension of the problem to the desired dimension $K$, while remaining in the feasible set of the problem. It is possible, for example, to set multiple roots in the directions of the users in adjacent cells. As will be shown in the simulations section, this will lead to extremely deep nulls in these directions.

There is a direct, but incomplete parallel with the rank-nullity theorem from linear algebra. Given constraints \eqref{eq:SOS}, we can construct the matrix $\mathbf A^H$, find an orthogonal projector to it, and use it as the basis $\mathbf Q$. Adding new, distinct directions to \eqref{eq:SOS} will reduce the dimension of the nullspace. However, addition of a multiple constraint will not restrict the nullspace of $\mathbf A^H$ further since its rank will not change. Thus, there are many distinct sub-faces which can be described using the relationship between varieties and ideals which are not readily described by way of the rank-nullity theorem.

\section{Generalization of  Generalized Sidelobe Canceller} 
In \cite{GriffithsJim}, Griffiths and Jim proposed an algorithm that has become the standard approach to linearly constrained adaptive beamforming. Their approach, referred to within \cite{GriffithsJim} as the GSC, uses a two step procedure in order to produce a beampattern with a fixed mainlobe and suppressed sidelobes. In the first step, a beampattern with a fixed response in the ``look'' direction is produced by convolving a vector of constraints with a normalized beamforming vector with the desired mainlobe response. In the second step, the signals in the ``look'' direction are blocked out, while the output power is minimized. 

If $y_{\mathrm w} (t)$ is the signal corresponding to the first part of the  Griffiths-Jim beamformer, and $y_{\mathrm n} (t)$ is the signal corresponding to the second part, then the overall beamformed signal is expressed as $y(t) = y_{\mathrm w} (t) - y_{\mathrm n} (t)$, where $t$ is discrete time index. In order to block the signal in the ``look'' direction, the authors use the assumption of ideal steering. To wit, they assume that the signal impinges on the broadside of the array. If we assume a ULA of $M$ antennas, the signal at the $m$-th antenna is $x_m(t) = s(t) + n_m(t)$. The assumption of ideal steering allows us to state that the desired signal $s(t)$ will be identical at each antenna (differing only by noise), and thus, a sufficient condition for blocking of the desired signal is $\mathbf w^T \mathbf{1} = 0$ where $\mathbf w$ is the blocking beamforming vector and $\mathbf{1}$ is the vector of all ones. 

Using the definition of the steering vector, it is can be seen that $\mathbf a(0) = \mathbf 1$, and thus, any beamforming vector satisfying $\mathbf w^T \mathbf 1 = 0$ will have a null at $\theta = 0$. Equivalently, $\mathbf w$ contains the coefficients of a polynomial with at least one root at $\alpha(0) = e^{j2\pi / \lambda_c d_x\sin(0)} = 1$. The $M - 1$ vectors $\mathbf w_m$ are compiled into an $(M-1) \times M$ matrix $\mathbf W_B$ with rows $\mathbf w_m^T$. It is clear that all $\mathbf w_m \in (x-1)\mathbb{C}_{M-1}[x]$, and thus, lie in the polynomial ideal $\mathcal{I} (1)$. 

The underlying algebraic structure allows several generalizing statements to be made. Instead of requiring the ideal steering, we can just require that $\mathbf w^T \mathbf a(\theta_0) = 0$, as $\mathbf w \in (x-\alpha(\theta_0))\mathbb{C}_{M-1}[x]$ is a necessary and sufficient condition for $s(k)$ to be blocked. Requiring that $\mathbf w$ be linearly independent for all $1\leq i \leq M-1$ implies that all the polynomials only share a single root at $\alpha(\theta_0)$. If multiple signals impinge upon the array from directions $\theta_l, \; 1\leq l \leq L$, and we wish to simultaneously block each of them in order to implement the GSC, the row-space of $\mathbf W_B \subset Q(x)\mathbb{C}_{M-2}[x]$ where $Q(x) = \prod_{l=1}^{L} (x-\alpha(\theta_l))$. In this case, we can have as many as $M - L$ vectors $\mathbf w_m$. In \cite{GriffithsJim}, the authors give only an example of the matrix $\mathbf W_B$ for $M = 4$. Moreover, as was shown in the previous subsection, the method proposed in this paper allows for further dimensionality reduction of the problem if computational complexity is an issue.

\section{Reformulation of Rank-Constrained SDP}
\label{eq:reformulation}
Given the construction of $\mathbf Q$ obtained, for example, as in Subsection~\ref{subsec:ConstQ}, it is now possible to reformulate the rank-constrained SDP such that it will be convex and thus solvable in polynomial time, yet have a solution which is rank-$K$ by construction. By introducing constraints of the form \eqref{eq:SOS} to the problem \eqref{eq:SDP_obj}--\eqref{eq:PSD} and using the findings of Section~\ref{sec:Null Shape}, the problem can be equivalently rewritten as
\begin{subequations}
	\begin{align}
	\min_{\mathbf X}\  &\; \mathrm{tr} \{ \mathbf X \mathbf Q^H \mathbf C \mathbf Q \} \label{eq:SDP_obj1c} \\
	\mathrm{s.t.}      &\; \mathrm{tr} \{ \mathbf X \mathbf Q^H \mathbf B_j \mathbf Q \} = \delta_j, \; j \in \mathfrak{J} \label{eq:affine1} \\
	&\; \mathbf X - \gamma \mathbf I \succeq 0 \label{eq:PSD2c},
	\end{align}
\end{subequations}
where $\gamma$ is some arbitrarily small positive real number and $\mathbf I$ denotes the identity matrix. 

From Theorem~1, it can be easily observed that the objective function \eqref{eq:SDP_obj1c} and the constraints \eqref{eq:affine1} are evaluated on a face of the PSD cone $\mathbb S_n^+$. As we have seen in Section~\ref{sec:Null Shape}, every point in the feasible set of this reformulated problem satisfies \eqref{eq:SOS} by construction. Moreover, the rank constraint has disappeared, and with it, the non-convexity of the problem. In addition, selecting a variety $\mathfrak{V}$ of correct size has resulted in a reduced problem dimension $K$. 

The constraint \eqref{eq:PSD2c} ensures that the solution to the optimization problem is full rank. Thus, the problem will have an optimal solution which is exactly of the desired rank $K$. If $\gamma$ is allowed to be $0$, the solution to the problem \eqref{eq:SDP_obj1c}--\eqref{eq:PSD2c}, $\mathbf X^{\star}$, may lie on the boundary of the cone $\mathbb{S}_k^+$ (note the dimension), and thus cause $\mathbf X^{\star}$ to drop rank. This serves as the main motivation for introducing the constraint \eqref{eq:PSD2c}. 

Moreover, the restriction has transformed the feasible set of the original problem \eqref{eq:opt2}--\eqref{eq:PSD1} to an affine slice of the positive definite cone. To wit, if there exists a feasible point to the problem at all, then the relative interior of the set is non-empty, and thus strong duality holds.

Using these observations, we can reformulate, for example, the transmit beamspace design problem for MIMO radar, that is, the problem \eqref{eq:opt2}--\eqref{eq:PSD1}. Introducing new matrices $\mathbf D \triangleq \mathbf Q^H \mathbf a(\theta) \mathbf a^H(\theta) \mathbf Q$ and $\mathbf H_i \triangleq \mathbf Q^H \mathbf B_i \mathbf Q$, and also using the cyclic property of the trace operator, the problem \eqref{eq:opt2}--\eqref{eq:PSD1} can be equivalently rewritten as
\begin{subequations}
	\begin{align}
	\min_{\mathbf X} \max_{\theta} &\; \left| G_{\mathrm d} (\theta) - {\mathrm tr} \{ \mathbf X \mathbf D \} \right| \label{eq:idealobj}  \\
	\mathrm{s.t.}      &\; {\mathrm tr}\{ \mathbf X \mathbf H_j \} = \frac{E}{N},\ j = 1,\cdots,N \label{eq:idealpow} \\
	&\; \mathbf X - \gamma \mathbf I
	\succeq 0 \label{eq:idealrank} .
	\end{align}
\end{subequations}

Finally, since the solution to the reformulated problem will be necessarily PD, as a consequence of the constraint \eqref{eq:idealrank}, it can be decomposed using the Cholesky decomposition as $\mathbf X^{\star} = \mathbf R \mathbf R^H$, giving us a simple way to recover the beamspace matrix $\mathbf W$ as $\mathbf W \triangleq \mathbf Q \mathbf R$. Obviously any unitary rotation of $\mathbf W$ recovered this way would also be a valid beamspace matrix.

\section{On the Conjugacy of Beamforming and Parameter Estimation}
It is well known that the beamforming and parameter estimation problems in array processing are conjugate with one another. In one, we are given the information of a target location and are tasked with fitting a beampattern fitting to this information in an optimal way. That is for minimum variance distortionless response beamforming, given the target location, and the signal auto-correlation matrix, minimize the variance of the array output while holding the distortionless response in the target location constant \cite{Capon}. In the other, we are given a signal auto-correlation matrix, and tasked with discovering the locations of the targets. In the case of transmit beamforming, the connection is also explicit. For transmit beamforming, the objective is to design a signal cross-correlation matrix given some information about the distribution of targets within an environment, while in parameter estimation, we are given a signal autocorrelation matrix, and asked to provide the target locations.

Consider $L$ sources impinging upon a ULA. 
The observation vector can be written as
\begin{align}
\mathbf x(t) = \mathbf A\mathbf s(t) + \mathbf n \nonumber
\end{align}
where $\mathbf A = [\mathbf a(\theta_1),\cdots,\mathbf a(\theta_L)]$, $\mathbf s(t)$ is the $L \times 1$ signal vector at time instant $t$ and $\mathbf n$ is observation noise. Assuming that $n \sim \mathcal{N}(0,\sigma^2 \mathbf I)$, and the number of snapshots $T$ is large enough, the sample covariance matrix of $\mathbf x(t)$ can be found as
\begin{align}
\mathbf R_{xx} = \frac{1}{T}\sum\limits_{t=1}^{T} \mathbf x (t) \mathbf x^H (t) 
= \mathbf A\mathbf S \mathbf A^H + \sigma^2 \mathbf I \nonumber 
\end{align}
If $T \geq N$ with probability equal to 1, the matrix $\mathbf R_{xx}$ is full rank. Since $\mathbf R_{xx}$ is Hermitian it has a full complement of real positive eigenvalues, and their associated eigenvectors must be mutually orthogonal. Thus, the sample covariance matrix can be written as
\begin{align}
\mathbf R_{xx} = \mathbf Q_{\mathrm s} \Lambda_{\mathrm s} \mathbf Q_{\mathrm s}^H + \mathbf Q_{\mathrm n} \Lambda_{\mathrm n} \mathbf Q_{\mathrm n}^H \nonumber
\end{align}
where $\mathbf Q_{\mathrm s}$, $\mathbf Q_{\mathrm n}$ are the matrices containing the signal and noise eigenvectors respectively, and $\Lambda_{\mathrm s}$, $\Lambda_{\mathrm n}$ are the diagonal matrices containing the signal and noise eigenvalues respectively. Since $\mathbf Q_{\mathrm n} \perp \mathbf Q_{\mathrm s}$, it follows that $\mathbf Q_{\mathrm n} \perp \mathbf A$, and thus, 
\begin{align}
\mathbf a^H(\theta_l) \mathbf Q_{\mathrm n} = 0,\ \forall \theta_l. \label{orth}
\end{align}

As has been shown in Section~\ref{sec:Null Shape}, the equation of the type \eqref{orth} can be satisfied if and only if the columns of $\mathbf Q_{\mathrm n}$, when viewed as the coefficients of polynomials with coefficients in $\mathbb{C}$, are in the univariate ideal generated by the variety given by the set of desirable roots $\mathfrak{V} = \{\alpha_1^*,\cdots,\alpha_L^*\}$. 

At once, the conjugacy between parameter estimation and beamforming problems become apparent in exact terms. In the transmit beamforming problem, we are given a variety, and design a signal cross-correlation matrix. While in the parameter estimation problem, we are given a sample covariance matrix and are asked to provide the variety which best explains it. Pisarenko's method, MUSIC, and root-MUSIC \cite{VanTrees} all use the orthogonality of the noise and signal subspaces. The root-MUSIC algorithm in brief is to find $\mathbf z$ such that 
\begin{align}
\mathbf z^H \mathbf Q_{\mathrm n} \mathbf Q_{\mathrm n}^H \mathbf z = 0 \label{eq:rootMUSIC}
\end{align}
which poses two fundamental challenges. The first one is thesignal/noise subspace selection, and the second is root selection \cite{MahdiMe}. 

\begin{figure*}[t]
	\centering
	\begin{minipage}[b]{0.48\textwidth}
		\begin{tikzpicture}[scale = 0.375]

\pgfplotsset{every axis legend/.append style={legend pos = north east},legend cell align = {left}}
        
\pgfplotsset{%
tick label style={font=\huge},
title style={font=\huge,align=center},
label style={font=\huge,align=center},
legend style={font=\huge}
            }

\begin{axis}[%
x label style={at={(axis description cs:0.5,-0.03)},anchor=north},
y label style={at={(axis description cs:-0.03,.5)},anchor=south},
x tick label style={rotate=45, anchor=north east, inner sep=0mm},
xlabel = {Angle $\theta$},
ylabel = {Signal Energy (dB)},
enlargelimits = true,
cycle list name = {color},
grid = major,
smooth,
scale = 3,
ymax = 30,
ymin = -70,
xmin = -76.25,
xmax = 76.25]

\addlegendentry{Proposed Method}

\addplot [ color = {blue}, 
           mark  = {asterisk},
           mark options = {solid},
           mark size = 4,
           style = {solid},
           line width = 1pt] table {datFiles/fig1-line1.dat};

\addlegendentry{ASL}

\addplot [ color = {red}, 
mark  = {none},
mark options = {solid},
style = {solid},
line width = 0.5pt] table {datFiles/fig1-line18.dat};

\addlegendentry{Roots}
\addplot [ color = {blue}, 
           mark  = {none},
           mark options = {solid},
           style = {solid},
           line width = 0.5pt] table {datFiles/fig1-line2.dat};

\addplot [ color = {blue}, 
           mark  = {none},
           mark options = {solid},
           style = {solid},
           line width = 0.5pt] table {datFiles/fig1-line3.dat};

\addplot [ color = {blue}, 
           mark  = {none},
           mark options = {solid},
           style = {solid},
           line width = 0.5pt] table {datFiles/fig1-line4.dat};

\addplot [ color = {blue}, 
           mark  = {none},
           mark options = {solid},
           style = {solid},
           line width = 0.5pt] table {datFiles/fig1-line5.dat};

\addplot [ color = {blue}, 
           mark  = {none},
           mark options = {solid},
           style = {solid},
           line width = 0.5pt] table {datFiles/fig1-line6.dat};

\addplot [ color = {blue}, 
           mark  = {none},
           mark options = {solid},
           style = {solid},
           line width = 0.5pt] table {datFiles/fig1-line7.dat};

\addplot [ color = {blue}, 
           mark  = {none},
           mark options = {solid},
           style = {solid},
           line width = 0.5pt] table {datFiles/fig1-line8.dat};

\addplot [ color = {blue}, 
           mark  = {none},
           mark options = {solid},
           style = {solid},
           line width = 0.5pt] table {datFiles/fig1-line9.dat};

\addplot [ color = {blue}, 
           mark  = {none},
           mark options = {solid},
           style = {solid},
           line width = 0.5pt] table {datFiles/fig1-line10.dat};

\addplot [ color = {blue}, 
           mark  = {none},
           mark options = {solid},
           style = {solid},
           line width = 0.5pt] table {datFiles/fig1-line11.dat};

\addplot [ color = {blue}, 
           mark  = {none},
           mark options = {solid},
           style = {solid},
           line width = 0.5pt] table {datFiles/fig1-line12.dat};

\addplot [ color = {blue}, 
           mark  = {none},
           mark options = {solid},
           style = {solid},
           line width = 0.5pt] table {datFiles/fig1-line13.dat};

\addplot [ color = {blue}, 
           mark  = {none},
           mark options = {solid},
           style = {solid},
           line width = 0.5pt] table {datFiles/fig1-line14.dat};

\addplot [ color = {blue}, 
           mark  = {none},
           mark options = {solid},
           style = {solid},
           line width = 0.5pt] table {datFiles/fig1-line15.dat};

\addplot [ color = {blue}, 
           mark  = {none},
           mark options = {solid},
           style = {solid},
           line width = 0.5pt] table {datFiles/fig1-line16.dat};

\addplot [ color = {blue}, 
           mark  = {none},
           mark options = {solid},
           style = {solid},
           line width = 0.5pt] table {datFiles/fig1-line17.dat};

\end{axis}
\end{tikzpicture}
		\vspace{-0.5cm}
		\caption{Solution to \eqref{eq:opt2}-\eqref{eq:PSD1}. An exact correspondence between $V$ and the beampattern nulls is observed.}
		\label{fig:Figure1}
	\end{minipage}
	\hspace{\stretch{2}}%
	\begin{minipage}[b]{0.48\textwidth}
		\begin{tikzpicture}[scale = 0.375]

\pgfplotsset{every axis legend/.append style={legend pos = north east},legend cell align = {left}}
        
\pgfplotsset{%
tick label style={font=\huge},
title style={font=\huge,align=center},
label style={font=\huge,align=center},
legend style={font=\huge}
            }

\begin{axis}[%
x label style={at={(axis description cs:0.5,-0.03)},anchor=north},
y label style={at={(axis description cs:-0.03,.5)},anchor=south},
x tick label style={rotate=45, anchor=north east, inner sep=0mm},
xlabel = {Angle $\theta$},
ylabel = {Signal Energy (dB)},
enlargelimits = true,
cycle list name = {color},
grid = major,
smooth,
scale = 3,
xmin = -76.25,
xmax = 76.25,]

\addlegendentry{SDR method}

\addplot [ color = {red}, 
           mark  = {none},
           mark options = {solid},
           style = {solid},
           line width = 1pt] table {datFiles/fig2-line1.dat};

\addlegendentry{ASL (SDR)}

\addplot [ color = {red}, 
           mark  = {none},
           mark options = {solid},
           style = {solid},
           line width = 1pt] table {datFiles/fig2-line2.dat};

\addlegendentry{Proposed Method}

\addplot [ color = {blue}, 
           mark  = {asterisk},
           mark options = {solid},
           mark size = 4,
           style = {solid},
           line width = 1pt] table {datFiles/fig2-line3.dat};

\addlegendentry{ASL (Proposed Method)}

\addplot [ color = {blue}, 
           mark  = {none},
           mark options = {solid},
           style = {solid},
           line width = 1pt] table {datFiles/fig2-line4.dat};

\end{axis}
\end{tikzpicture}			
		\vspace{-0.5cm}
		\caption{Solution to \eqref{eq:opt2}-\eqref{eq:PSD1}. A gap of 19.8 dB in terms of ASL for the proposed and SDR-based methods while a very close agreement between the passbands is observed.}		
		\label{fig:Figure2}
	\end{minipage}
	\vspace{-0.3cm}
\end{figure*}

Subspace selection is typically performed by selecting the eigenvectors which correspond to the $N-L$ eigenvalues. This works very well in practice in cases where there are many samples $T$, and a high signal-to-noise ratio. However, if either of these assumptions is invalid, a phenomenon when some columns of $\mathbf Q_{\mathrm s}$ will be erroneously selected to form $\tilde{\mathbf Q}_{\mathrm n}$ may occur. 

Since \eqref{eq:rootMUSIC} is a sum of sum of squares polynomials, it can only be zero if each of the sum of squares polynomials is zero simultaneously, or equivalently, if each column of $\mathbf Q_{\mathrm n}$ is in the polynomial ideal of the variety $\mathcal{I} (\mathfrak{V})$. This provides a new criterion for the selection of noise eigenvectors from $\mathbf Q$: choose the eigenvectors which are ``closest''(with respect to some appropriate measure) to being in a univariate ideal. 

Let $f_1,\cdots,f_N$ be the polynomials with coefficients equal to the entries of the eigenvalues of $\mathbf Q$, and assume that they are ordered such that $f_1,\cdots,f_L$ correspond to the signal eigenvectors, and $f_{L+1},\cdots,f_N$ correspond to the noise eigenvectors. In the absence of noise, the polynomial ideal structure enforces that
\begin{align}
f_{L+1} &= gh_1 \nonumber \\
f_{L+2} &= gh_2 \nonumber \\
&\vdots \nonumber \\
f_N &= gh_{N-L} \nonumber
\end{align}
where $\mathbf g \triangleq \prod\limits_{l = 1}^{L} (x-\alpha_l^*)$ and $h_1,\cdots,h_{N-L}$ are coprime polynomials (due to the mutual orthogonality of the eigenvectors, these polynomials must necessarily be coprime). 

In the presence of noise, however, $g$ becomes perturbed and the system of equations becomes
\begin{align}
f_{L+1} &= \tilde{g}_1h_1 \nonumber \\
f_{L+2} &= \tilde{g}_2h_2 \nonumber \\
&\vdots \nonumber \\
f_{N} &= \tilde{g}_{N-L}h_{N-L} \nonumber 
\end{align}
where $\tilde{g}_i \triangleq \prod\limits_{l = 1}^{L} (x - \alpha_l^* + \epsilon_{l,i})$.
 
Naturally, the identification of $g$ solves both the subspace selection problem and the root selection problem. The root-MUSIC polynomial therefore, in actuality, has no roots, while the ideal root-MUSIC polynomial has only $L$ roots which correspond exactly to the target locations. What is meant by the number of roots in the classical root-MUSIC algorithm is the sum of the number of roots of each individual sum of squares polynomial. However, due to the polynomial ideal structure, $L(N-L)$ of these roots are very close to one another. Thus, instead of choosing the roots which are closest to the unit circle, it makes sense to choose the roots which are closest to each other.

\label{sec:simulation}
\section{Simulation Results}
To exhibit the capabilities of the proposed method, we present two distinct simulation examples based on solving the problem \eqref{eq:opt2}--\eqref{eq:PSD1} in two different scenarios. In both scenarios, the underlying system setup remains the same: a ULA with $N = 20$ antenna elements spaced at multiples of $\lambda_c/2$ acts as a transmitter with the goal of closely matching a desired radiation pattern. As this is a transmit beamforming problem, noise is not present in the model. Array imperfections are not considered in the simulations. In both scenarios, the method based on SDR is offered as a comparison. Throughout all simulations, a transition region of $5^o$ is allowed on either side of the passband region.

{\bf Example 1:} In this example, a uniform prior target distribution over the interval $\Theta = [-15^o,15^o]$ is assumed. Thus, the goal is to transmit energy uniformly within the sector $\Theta$ while suppressing transmitted energy as low as possible elsewhere. To design $\mathbf Q$, null directions are selected outside of the sector $\Theta$ in a roughly uniform spread. The number of transmitted waveforms is $K = 4$, and thus, the number of nulls to be set is $L = N - K = 16$. The variety chosen is $\mathfrak{V} = \{\pm \alpha(75^o),\pm \alpha(60^o),\pm \alpha(50^o),\pm \alpha(43^o),\pm \alpha(34^o),\pm \alpha(33^o),\\ \pm \alpha(26^o),\pm \alpha(22^o)\}^*$. Given the variety $\mathfrak{V}$, the matrix $\mathbf Q$ is constructed in line with the findings of Section~\ref{sec:Null Shape}. After that, the problem \eqref{eq:idealobj}--\eqref{eq:idealrank} is solved. 

\begin{figure*}[t]
	\centering
	\begin{minipage}[b]{0.48\textwidth}
			\begin{tikzpicture}[scale = 0.375]

\pgfplotsset{every axis legend/.append style={legend pos = south east},legend cell align = {left}}
        
\pgfplotsset{%
tick label style={font=\huge},
title style={font=\huge,align=center},
label style={font=\huge,align=center},
legend style={font=\huge}
            }

\begin{semilogyaxis}[%
x label style={at={(axis description cs:0.5,-0.03)},anchor=north},
y label style={at={(axis description cs:-0.03,.5)},anchor=south},
x tick label style={rotate=45, anchor=north east, inner sep=0mm},
xlabel = {Angle $\theta$},
ylabel = {Signal Energy (dB)},
enlargelimits = true,
cycle list name = {color},
grid = major,
smooth,
scale = 3]

\addlegendentry{SDR method}

\addplot [ color = {blue}, 
           mark  = {asterisk},
           mark options = {solid},
           style = {solid},
           line width = 1pt] table {datFiles/fig11-line1.dat};

\addlegendentry{Proposed Method}

\addplot [ color = {red}, 
           mark  = {*},
           mark options = {solid},
           style = {dashdotted},
           line width = 1pt] table {datFiles/fig11-line2.dat};

\end{semilogyaxis}
\end{tikzpicture}
			\vspace{-0.5cm}
			\label{fig:Figure11}
			\caption{Solution to the problem \eqref{eq:opt2}-\eqref{eq:PSD1}. Both methods tested performe almost exactly the same in terms of passband ripple, null-depth, and roll-off.}
	\end{minipage}	
	\hspace{\stretch{2}}%
	\begin{minipage}[b]{0.48\textwidth}
	\begin{tikzpicture}[scale = 0.375]

\pgfplotsset{every axis legend/.append style={legend pos = south east},legend cell align = {left}}
        
\pgfplotsset{%
tick label style={font=\huge},
title style={font=\normalsize,align=center},
label style={font=\huge,align=center},
legend style={font=\huge}
            }

\begin{semilogyaxis}[%
x label style={at={(axis description cs:0.5,-0.03)},anchor=north},
y label style={at={(axis description cs:-0.03,.5)},anchor=south},
x tick label style={rotate=45, anchor=north east, inner sep=0mm},
xlabel = {Angle $\theta$},
ylabel = {Signal Energy (dB)},
enlargelimits = true,
cycle list name = {color},
grid = major,
smooth,
scale = 3]

\addlegendentry{SDR method}

\addplot [ color = {blue}, 
           mark  = {asterisk},
           mark options = {solid},
           style = {solid},
           line width = 1pt] table {datFiles/fig12-line1.dat};

\addlegendentry{Proposed Method}

\addplot [ color = {red}, 
           mark  = {*},
           mark options = {solid},
           style = {dashdotted},
           line width = 1pt] table {datFiles/fig12-line2.dat};

\end{semilogyaxis}
\end{tikzpicture}
	\vspace{-0.5cm}
	\label{fig:Figure12}
	\caption{Solution to the problem \eqref{eq:opt2}-\eqref{eq:PSD1}. The null-depth for the proposed method is several orders of magnitude deeper than that of the SDR-based method.}
	\end{minipage}
	\vspace{-0.3cm}
\end{figure*}

Fig.~2 shows the solution of \eqref{eq:idealobj}--\eqref{eq:idealrank} under aforementioned system assumptions. The vertical lines describe the locations of the nulls corresponding to the set $\mathfrak{V}$. It is observed that the correspondence for the proposed method is exact, and ASL stands here for average sidelobe level, which is an average level of sidelobes in all directions outside of the sector of interest obtained by the proposed method.

A comparison between the proposed method and the solution obtained based on SDR approach to the problem \eqref{eq:idealobj}--\eqref{eq:idealrank} is shown in Fig.~3. The ASL corresponding to the solution of the problem \eqref{eq:opt2}--\eqref{eq:PSD1} obtained via SDR is -4.73~dB. A gap of 19.8~dB in terms of ASL while a very close agreement between the passbands is observed. Indeed, the peak sidelobe level of the proposed method is 10~dB below the ASL of the SDR result, while the mean-squared error (MSE) between the solution of the proposed method and the desired beampattern is $229$, compared to $247$ in the case of the SDR-based method. 

It should be noted that a rank-$K$ approximation of $\mathbf X_{SDR}^\star$ still has to be achieved. However, $\mathbf X_{SDR}^\star$ is typically extremely ill-conditioned ($\kappa \approx 10^{10}$), and thus, an exact reconstruction of $\mathbf X_{SDR}^\star$ using the full rank matrix decomposition is not possible. If constraints of the type \eqref{eq:idealrank} are used to restrict the condition number of $\mathbf X$, they will be active, and the solution of $\mathbf X_{SDR}^\star$ will thus degrade further. Lastly, there are no optimality guarantees for randomization techniques for the case of general rank approximations. The comparison provided therefore represents the most favorable comparison possible with the SDR-based method, yet the results of the proposed method are dramatically better.

{\bf Example 2:} In this scenario, the situation is somewhat reversed to the previous example. Instead of a uniform prior distribution of targets, we are given a specific location in which we must not transmit energy, and no other prior information about the environment. In such a case, it is most sensible to uniformly transmit energy in all directions except for the given one. In this scenario, the direction in which we wish to transmit no energy is $\theta = -13^o$. 
To design $\mathbf Q$, the variety is chosen to be the singleton $\mathfrak{V} = \alpha(-13^o)$. Note that in this case, $\mathbf Q$ corresponds exactly to the blocking matrix in Griffiths-Jim beamformer without the assumption of ideal steering, for a source impinging from a direction of $-13^o$. 

Fig.~4 shows the solution of the problem \eqref{eq:opt2}--\eqref{eq:PSD1} in comparison to the one provided by the SDR-based method. It can be seen that the methods perform almost exactly the same in terms of passband ripple, null-depth, and roll-off.

Given the findings of Subsection~\ref{sec:dimreduce}, it is important to investigate both the capability of the proposed method to perform the dimensionality reduction given only a single null-direction, and the effect that the dimensionality reduction has on the performance of the method. To test this capability, we set the variety by which $\mathbf Q$ will be designed to be $\mathfrak{V}' = \{\alpha(-13^o),\alpha(-13^o),\alpha(-13^o)\}$. Since $\mathfrak{V}' \supset \mathfrak{V}$, we know that $\mathcal{I}(\mathfrak{V}) \supset \mathcal{I}(\mathfrak{V}')$, and thus, $\mathcal{C}(\mathbf Q) \supset \mathcal{C}(\mathbf Q)$, which in turns means that the restriction is contained by the feasible set as the original problem \eqref{eq:opt2}--\eqref{eq:PSD1}. 

Fig.~5 exhibits a performance comparison between the proposed and SDR-based methods. It is observed that the null-depth for the proposed method is several orders of magnitude deeper than that of the SDR-based method. As the proposed method only uses $17$ degrees of freedom, compared to the $20$ used by the SDR-based method, pass-band ripple increases. In the case of the Griffiths-Jim beamformer of \cite{GriffithsJim}, this would correspond to a lessened ability to minimize the noise variance.

\section{Conclusions}
\label{sec:conclusions}
A new approach to solving a class of the rank-constrained SDP problems has been presented. Instead of relaxing such non-convex problem to a feasible set of positive semidefinite matrices, we restrict the problem to a space of polynomials whose dimension is equal to the desired rank of the solution. The resulting optimization problem is then convex and can be efficiently and exactly solved, while the solution of the original rank-constrained SDP problem can be exactly recovered from the solution of the restricted one through a simple matrix decomposition. We show how this approach can be applied to solving some important signal processing problems with so-called null-shaping constraints, which enforce the desired algebraic structure to the feasible set. Specifically, we have shown how to apply the proposed approach to address such signal processing problems as transmit beamspace design in MIMO radar, downlink beamforming design in MIMO communication, and GSC design. As the beamforming and parameter estimation problems are known to be conjugate of each other, we have also shown, as a byproduct of the main studies here, the conjugacy in exact terms, and have formulated a new exact algebraic theoretically motivated criterion for signal/noise subspace identification. Simulation results performed for the problem of rank-constrained beamforming design have shown an exact agreement of the solution with the proposed algebraic structure, as well as significant performance improvements compared to the existing SDR-based method.      

\end{document}